# Quantifying a certain "advantage law": minority game with above the rules agents

J.R.L de Almeida[1] and J. Menche[1,2]

[1]Departamento de Física, Universidade Federal de Pernambuco, 50670-901, Recife, PE, Brasil
almeida@df.ufpe.br
[2]Fakultät für Physik und Geowissenschaften, Universität Leipzig, Linnéstraße 5, 04103 Leipzig, Germany
joergmenche@gmx.de

## ABSTRACT

In this work the properties of minority games containing agents which try to winning all the time are studied by means of computational simulations. We have considered several ways of introducing above the rules clever players using "strategies" which try to outdo the others endowed with statistically equivalent strategies and compared the resulting behaviours of the ensemble. It is shown that by introducing such agents the overall performance of the system gets significantly poorer. While the introduction of a very small fraction of these never-loosing-players may not destroys the unordered / ordered phase transition of the standard minority game we find that even for a low concentration of their presence only a state "worse" than random coin toss choices sets in. These special agents/players have the role of impurities or vacancies in spin systems and their presence may lead to a critical concentration where the usual phases are washed out.

The minority game (MG) is a very simple model of interacting agents. Nevertheless it shows fascinating properties and poses novel challenging questions to statistical mechanics and its methods and ideas. It is a very simple toy model for complex systems and may come to play the same role in science as the Ising model, applicable in many interdisciplinary subjects from physics to economy and biology to name a few. In short, its main idea is the following: an odd number of selfish players repeatedly has to choose one out of two possible actions. These could be for example buying or selling at the stock market, taking the bus or the metro, which internet router is the fastest (ie. less crowded). As it will become clearer in the examples, the players that are in the minority side will win the game. If there are more sellers than buyers, the price will fall and the buyers will be in advantage. If there are lots of people in the metro it would have been much more comfortable to take the bus and so on. This kind of game was inspired by Arthur's ``El Farol'' problem [1]. There it was an overcrowded bar that rose his interest of how a population of heterogeneous players interacts.

In this section we give a short and necessarily incomplete summary of the MG and their well-known properties. Then we shall present our modified models and compare them with the standard model. In the MG that was introduced in [2,3], $N$ agents (or players) choose at each time step $\tau$ one out of two possible options, say 0 and 1. Agents that belong to the smaller of both groups win. They act completely independent, without any possibility of communication or interaction. Their only available information about the other player's actions is a ``history'' (or ``memory'') of the last $m$ games' winners. There are two possible values for each of the $m$ entries of the history, so there are $2^m$ possible different histories. Every player is equipped with a set of $S$ strategies to determine their next choice. As the decision depends on the current history of the last games, these strategies contain $2^m$ values, each of them representing the choice for one of the possible histories. All of the $2^m$ entries can be 0 or 1, so there are $2^{2^m}$ different strategies. After all agents having made their choice and the winning group being determined, the agents compare the predictions that their strategies made with the present outcome. If a strategy was right, it will be rewarded with one point. The agents will always use the best strategy, that means the one with the most points, to take their choice. An interesting quantity in the MG is the standard deviation

$$\sigma = \sqrt{\frac{1}{R}\sum_{i=0}^{R}\left(N_i^{(1)} - \tfrac{N}{2}\right)^2}$$

where R is the total number of rounds, $N_i^{(1)}$ is the number of agents playing 1 (or 0) in round *i*. The standard deviation is a measure for the efficiency of the system at distributing the limited resources. $\frac{N}{2}(\approx \frac{N-1}{2})$ is the maximum number of points to be won per time step. Small variations around this value mean therefore, that the system is very efficient. Fig.1 shows our results for two different values of *N*.

Numerical simulations as fig.1 or in [4], as well as analytical calculations [5] show that the system undergoes a phase transition. The two phases are separated by the minimum of the curve. On the left side of the minimum we observe that the value of σ varies widely from the medium value, whereas on the right side these variations vanish almost completely. In [6] it is shown that all the curves for different values of *N* and m fall on a universal curve by using a $\sigma^2/N$ vs. $2^m/N$ plot, see fig.2. For large values of the ratio $\alpha = 2^m/N$ the system displays a cooperative phase ($\alpha > \alpha_c \approx 0.34$), for small values ($\alpha < \alpha_c$) the coordination of the agents is worse than the one of a ``random system``. The ``random system" consists of agents that make their choice by coin toss, without using any strategy.

These general properties do not depend on the number of available strategies *S* that the agents can use. This is shown in fig.3 for S=4.
In fig.4 we directly compare the performances of two ensembles with S=2 and S=4. There we see that four strategies are less efficient than two strategies, as it is explained in [6]. Only for larger values of *m* (almost random choice) the results are the same.

Among many of possible generalizations the standard minority game with two choices can be generalized to more than just two options [7]. In fact life would be kind of sad, if there always were only two options to choose from. The qualitative behaviour remains the same as it is for two choices. Only the value $\alpha_c$ where the phase transition occurs depends on the number of choices. Here we undertake a generalization which focuses on the effects on the system of interacting heterogeneous agents when privileged agents take out part of the generated resources available to the "honest" players. There are many ways of trying to do so and in the next section we present some of them.and its effects, notably the suppression of the ordered / disordered phase transition if the population of the above-the-rules players is higher than a given threshold. In terms of spins systems [6], these agents play the role of impurities or vacancies and their presence naturally lead to a concentration threshold or impurity concentration limits where order / disorder phase transition is washed out.

So agents which get above the average rewards no matter the global state of the system may seems a common place . The method for achieving this goal may vary from naturally inherited better strategies to breaking the rules. We wanted to introduce this idea in the minority game, specially the latter case, and analysed several different ways of doing so:

**M1:** in addition to N agents equipped with S normal strategies like in the standard MG there are k agents which always join the winning (minority) side thus sharing the rewards. A penalty is introduced : if the total number of the agents sharing the points in a run exceeds the available resources, they become loosers.

In figure 4 it is shown the standard deviation vs. memory for this model for k=0, 4, 6, and 10 which for the number of agents considered is approximately the percentage of "clever" agents in the system. Clearly the overall performance of the adaptive system becomes worse as soon as k is different from zero. By depriving the normal players of their reward the extra agents increase the fluctuations in the number of agents in the minority side thus favoring higher fluctuations and this happens for any concentration of the extra players for model M1. At about 10% of their presence the averaged standard fluctuation is always above the random coin tossing case (horizontal line in the graph) and there is no longer a optimum distribution (minima for sigma).

**M2:** in addition to agents equipped with S normal strategies like in the standard MG there are k agents which always join the winning (minority) side sharing evenly with them the rewards for that run.

In figure 5 it is shown the standard deviation vs. memory for this model for k=0, 5,10 and 20. Clearly the overall performance of the adaptive system is again worse as soon as k is different from zero and again depriving the normal players of their full reward, the extra agents increase the fluctuations in the number of agents in the minority side thus favouring higher fluctuations ,i.e., a large spread in σ, and this happens for any concentration of the extra players for model M2. In this case for low memories their presence may improves somewhat the performance (which always a bad one) but at high memory values for all k considered the system performs worse than the random coin case.

In figure 6 it is shown the utility function ( a measure of the effectiveness of the system in distributing resources [9] ) for the above two models. Again, both models have their performance decreased when compared with the standard MG model.

**Metc..:** in addition to agents equipped with S normal strategies like in the standard MG there are k agents which always join the winning (minority) side and they exclude for rewarding the k poorest performers up to that time among the winners; or they look like normal agents if their strategies win rewards otherwise they change the reward assignment; or etc… There is an infinity of ways of implementing above-the-rules agents as well as complexing the model by introducing evolutionary agents and penalties. However, the above two simple models should illustrate the point that in these adaptive systems agents breaking the rules of the majority do so at the expenses of decreasing the overall performance.

In conclusion, the effect of introducing agents which somehow gets a share of the resources available, no matter how, seems to worsen the system resources distribution and effectiveness. Looking around, most of the time, this should come with no surprise. What may be wonderful is that by studying a very simple toy model, amenable to analytical study [5,6], it arises the possibility of quantifying the effects of such agents either by numerical simulations (experiments) or analytically which we leave for future works.

## Acknowledgments

J.Menche wishes to thank the DAAD (German academic exchange service) . J.R.L. de Almeida would like to thank the DAAD too and all those who inspired this work ( the so called "advantage law" is a popular brazilian saying applied whenever someone tries to overcome others at the expenses of their rights. A simple example: overtaking others in a queue).

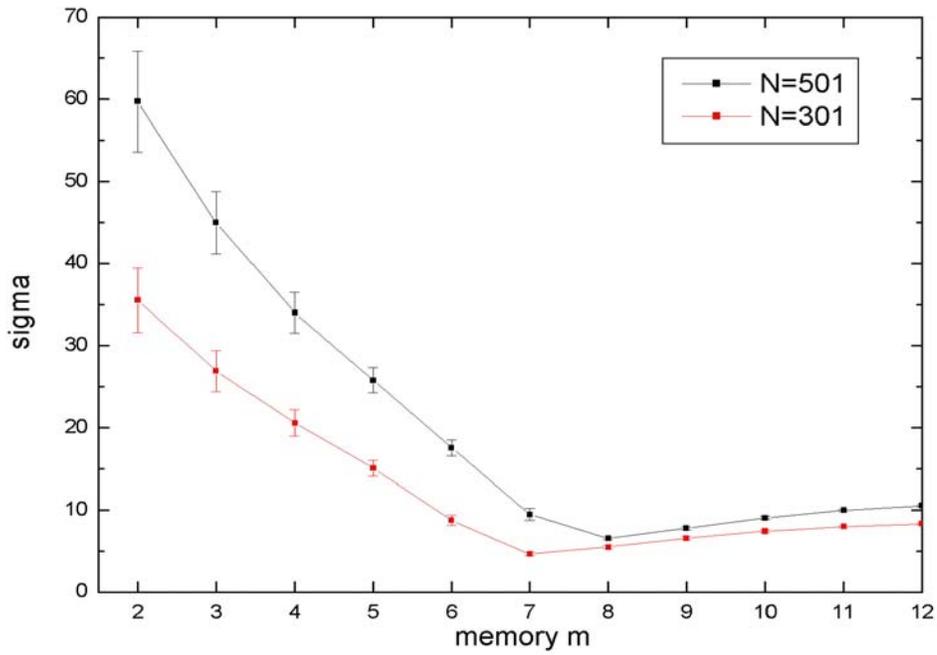

FIG. 1: SMG; $\sigma$ vs m ; S = 2; averages over 100 realizations

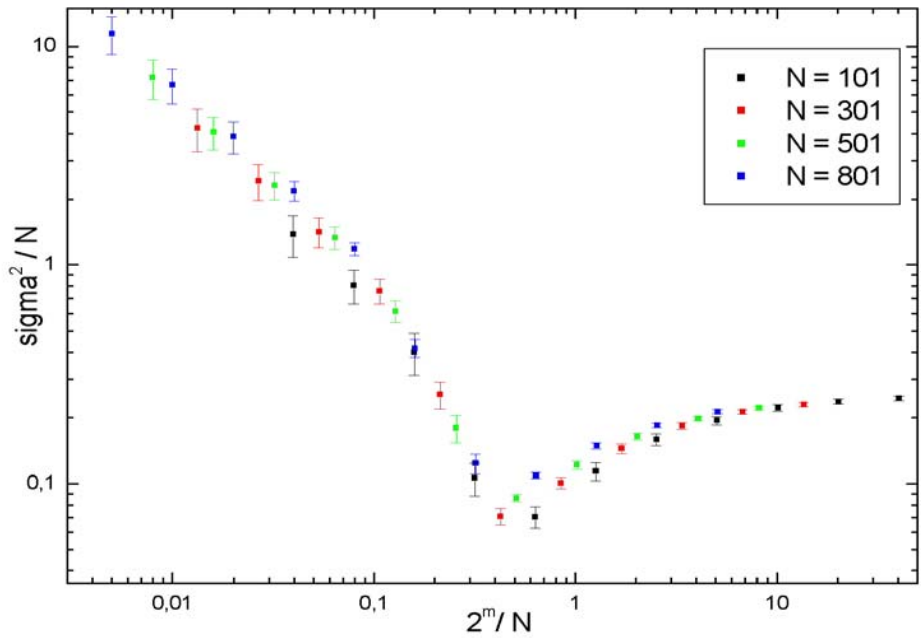

FIG. 2: SMG; $\sigma^2 / N$ vs $2^m / N$ ; S = 2; averages over 100 realizations

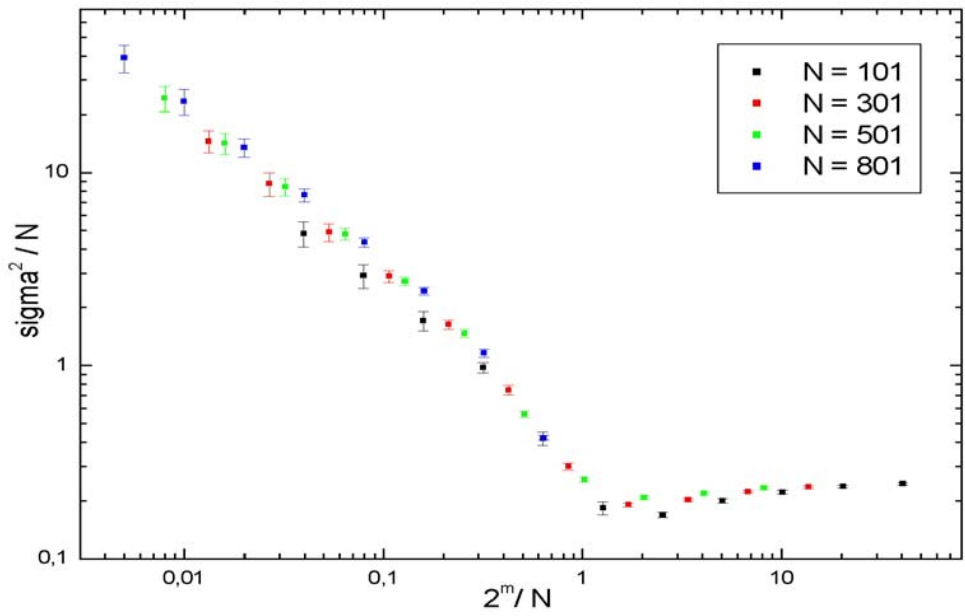

FIG. 3: SMG; $\sigma^2 / N$ vs $2^m / N$ ; S = 4 averages over 100 realizations;

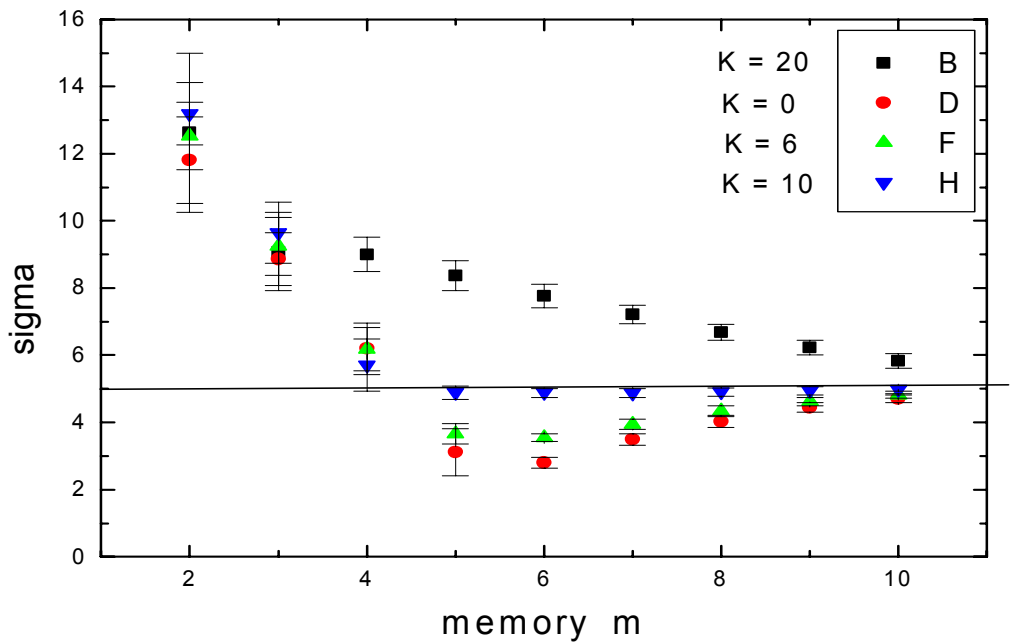

FIG. 4: MG; $\sigma$ vs m ; N=101 , S = 2; model M1; averages over 200 realizations

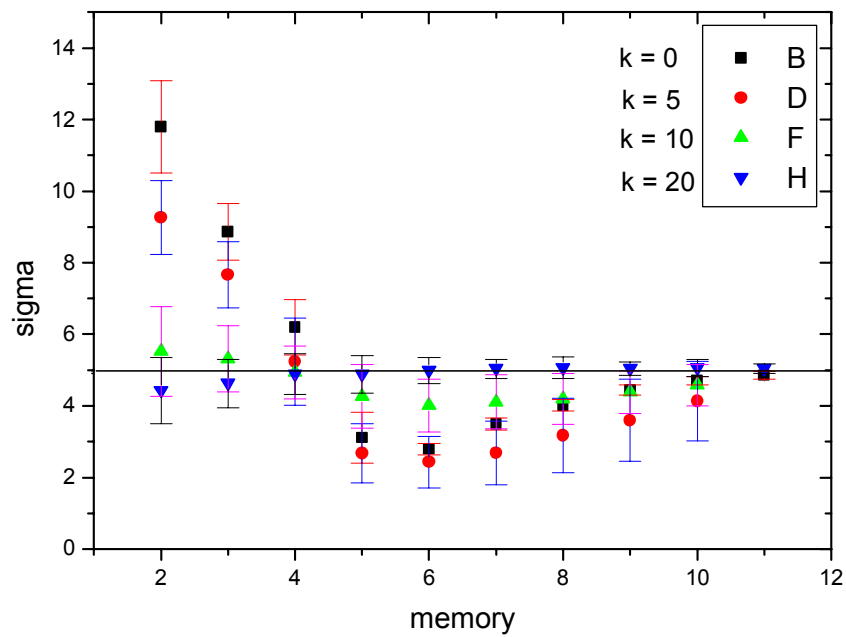

FIG. 5: MG; σ vs m ; N=101 , S = 2; model M2; averages over 200 realizations

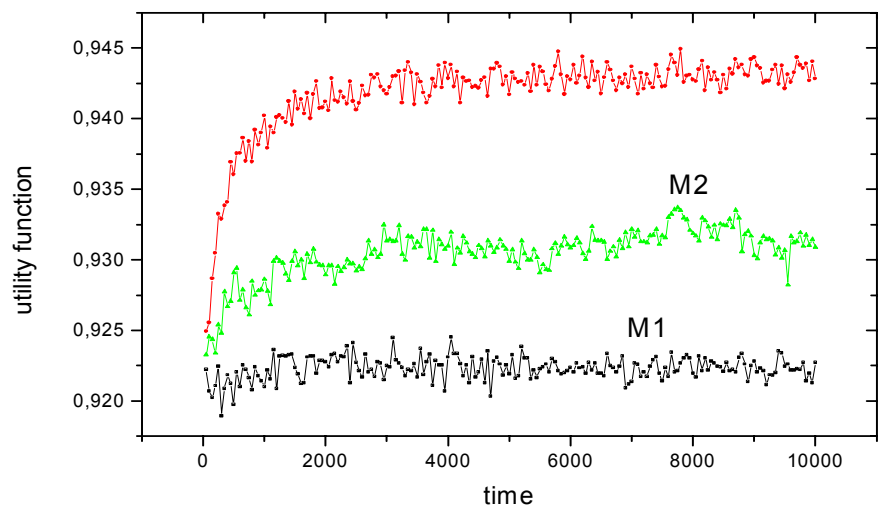

FIG. 6: MG; utility function vs time; N=101 , S = 2, m=8; averages over 200 realizations